# Mandating Code Disclosure is Unnecessary – Strict Model Verification Does Not Require Accessing Original Computer Code

**Sasanka Sekhar Chanda**
**Professor, Strategic Management,**
**Indian Institute of Management Indore**
sasanka2012@gmail.com

**Abstract.** Mandating public availability of computer code underlying computational simulation modeling research ends up doing a disservice to the cause of model verification when inconsistencies between the specifications in the publication text and specifications in the computer code go unchallenged. Conversely, a model is verified when an independent researcher undertakes the set of mental processing tasks necessary to convert natural language specifications in a publication text into computer code instructions that produce numerical or graphical outputs identical to the outputs found in the original publication. The effort towards obtaining convergence with the numerical or graphical outputs directs intensive consideration of the publication text. The original computer code has little role to play in determining the verification status—verified/ failed verification. An insight is obtained that skillful deployment of human intelligence is feasible when effort-directing feedback processes are in place to appropriately go around the human frailty of giving up in the absence of actionable feedback. This principle can be put to use to develop better organizational configurations in business, government and society.

---

## 1 Introduction

In agent-based modeling research, a set of theorized mechanisms are translated into computer instructions in order to elicit macro-level outcomes from stylized micro-level dynamic actions of agents (de Marchi and Page 2009). The aim of **model verification** is to confirm that the specifications of the agent model given in the publication text are faithfully translated into computer instructions.[1] A published computational simulation model (e.g. an agent-based-model – ABM) fails verification if (a) the original computer model fails to incorporate one or more specification elements given in the publication text and/or (b) the original computer code model contains logic for which specifications are absent in the publication text and/or (c) the original computer code model implements a particular, hard-to-fathom meaning for an ambiguous statement in the publication text. Absent appropriate model verification, there is always an element of doubt regarding model findings. This constrains the use of model results in subsequent research. Moreover, researchers also find it difficult to extend the model to develop new knowledge.

---
[1] Interested readers may refer to the Appendix for some examples of model verification studies.

In order to promote transparency in computational simulation modeling research, commentators in *Nature* (e.g. Ince *et al.* 2012) and elsewhere (e.g. Donoho 2010) argue that journals should mandate public availability of the computer code as a pre-condition for publishing ABM research. The commentators appear to assume that public scrutiny of the code is *sufficient* to ensure that research is not compromised. I contest this view. I contend that reliance on unproven virtues of public scrutiny is misplaced. The need of the hour is to develop understanding regarding what it takes to validate a computational simulation model / ABM. Only then it makes sense to infer what policies journal editors may put in place to encourage model verification.

## 2      Model verification must involve conceptual replication

The central thesis of this article is that, although a demand is frequently made that authors publishing computational simulation modeling work make their computer code publicly available, there is inadequate appreciation as to how such public availability helps or hinders **model verification**. The popular belief (Donoho 2010; Ince *et al.* 2012) appears to be that the very prospect of having their code open for scrutiny will make authors doubly cautious about the quality of their code, and compel them to maintain a close synch between the conceptual specification—as described in the publication text—and the actual implementation in the computer code. This is wishful thinking. If a computer code contains invalid assumptions or approximations, probably the only way those come to light is when an independent researcher (or an independent research team) attempts to develop computer code solely from the specifications in the publication text *without looking at the original code* (Chanda and Miller 2019; Dent 2012; Wilensky and Rand 2007). This process is referred to as *conceptual replication* (Chanda and Miller 2019). The crux of model verification lies in ensuring that the set of natural language statements in a publication text suggest a logically equivalent set of computer instructions, *across researchers*.

Variances between the graphical/numerical output produced by the independently developed code and the graphical output presented in the publication text serve to indicate that the conceptual model in the publication text and the original computer code implementation are not in synch. Subsequent comparison of the original and independently developed computer codes can identify the actual deviations between the conceptual model implied in the publication text and the model implemented in the original code.

## 3 Formal Demonstration

In this section, I formally demonstrate that model verification can be appropriately accomplished without having to consult the computer code of the original/ pioneering publication. Moreover, I show that the alternative comprising comparing the implications of the pioneer's computer code with the statements in the publication is an inferior and error-prone approach.

### 3.1 Terminology and Relationships

   I. Let **C** represent the conceptual specification in the text of a computational simulation (agent-based) model publication that is taken up for model verification.
  II. Let **I** represent the computer code underlying **C**, as designed by the pioneering authors who first publish the agent-based-modeling paper.
 III. Let **G** represent the set of graphs (or numerical output values) reported in the original publication. Presumably, **G** materialized by executing the computer code **I**.
  IV. Let **I\*** represent computer code developed by referring only to **C** and not at all referring to **I**.
   V. Let **G\*** represent the set of graphs (or numerical output values) obtained by executing the computer code **I\***.
  VI. Let **C\*** be the equivalent set of implied natural language statements based on the implemented model in the pioneer's computer code **I**. **C\*** is developed by appropriately summarizing the logic in the pioneer's code **I**, and expressed by a set of natural language (English) statements.

Above definitions entail following relationships.

   **(a)** From [I], [II] and [III] it follows that the conceptual specification in the text of an agent-based model publication (**C**) got translated into **I**—computer code mapping specification statements in **C** to implementation in code logic. Further, execution of the computer code **I** produces the graphical (or numerical) output **G**.

   $$\mathbf{C} \rightarrow \mathbf{I} \rightarrow \mathbf{G} \qquad \ldots [1]$$

   Above relationship is to be read as follows: from **C** we obtain **I**, and from **I** we obtain **G**. We may note that the chances of any problematic issue in the second link $\mathbf{I} \rightarrow \mathbf{G}$ is very low, unless certain operating system (OS) internals (like handling of floating point calculation and/or random number generation) are vastly different and simulation results change upon change in design of such OS internals. Model verification, therefore, entails verifying that the first link **C → I** is error-free. We shall see subsequently that this can be done in two ways. The first approach involves starting from **I**, deriving a **C\*** from **I** (as per [VI]) and comparing **C** and **C\***. The second approach, conceptual replication, involves developing code **I\*** from **C** (as per [IV]) and then comparing the output graphs/ data tables from **G\*** obtained by executing **I\*** (as per [V]) with the data tables/ graphs **G** provided in the publication text.

(b) From [VI], that we obtain **C\*** from **I** we have:

$$\mathbf{I} \to \mathbf{C^*} \qquad \ldots [2]$$

If we attempt model verification by the first approach we need to check the extent of conformance between the computer code model specification statements in **C** with **C\***. We wish to judge whether the specifications in **C** and **C\*** are sufficiently similar. I use the "≈" sign to denote sufficient similarity and write the next relationship in form of a question, by adding a "?" sign at the end.

$$\mathbf{C^*} \approx \mathbf{C}? \qquad \ldots [3]$$

(c) If we attempt model verification by the second approach—conceptual replication—from [IV] and [V] we have:

$$\mathbf{C} \to \mathbf{I^*} \to \mathbf{G^*} \qquad \ldots [4]$$

Again, we may note that the link **I\* → G\*** is likely to be straight-forward as in the previous case. The question at the very heart of model verification constitutes the fidelity of the (other) link **C → I\*** since it sets out to inquire whether a researcher can independently arrive at same (or sufficiently similar) conclusions as the pioneer, by following the specifications articulated in the publication text. Such is done by checking whether **G** and **G\*** are sufficiently similar. I use the "≈" sign to denote sufficient similarity and write the next relationship in form of a question, by adding a "?" sign at the end.

$$\mathbf{G^*} \approx \mathbf{G}? \qquad \ldots [5]$$

By incorporating above relationships, in **Figure 1**, I present a schematic diagram for the first approach for model verification—model verification using the computer code of the pioneer's publication. In **Figure 2** in the following section, I present a schematic diagram for the second approach for model verification—conceptual replication (i.e. model verification *without* referring to the pioneer's computer code).

**Figure 1. Model verification using the computer code of the pioneer's publication**

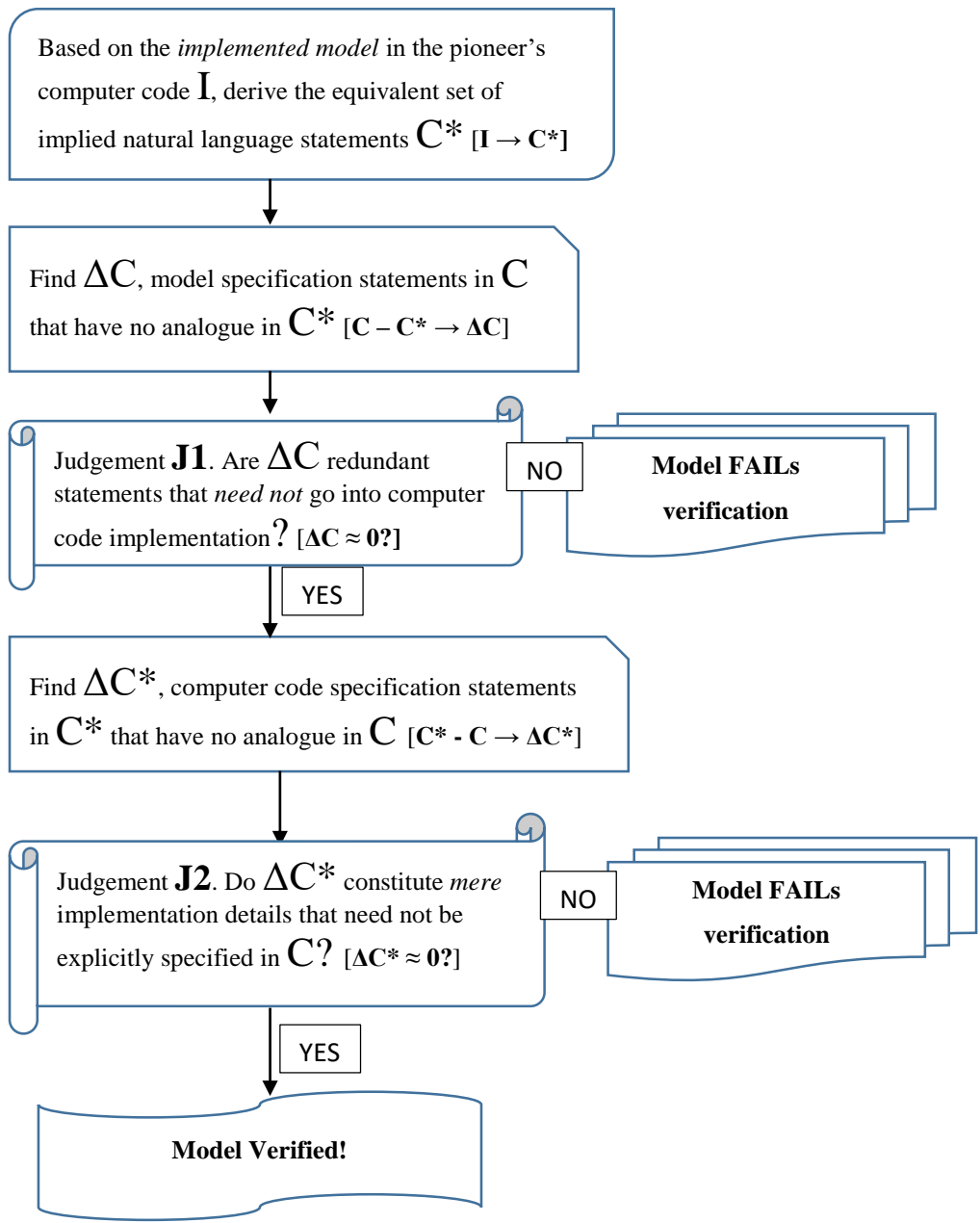

*Note.* This method uses the pioneer's publication text **C** and assumes that the pioneer's code **I** is available. This does not make use of the graphs / output data tables **G** provided in the pioneer's publication.

## 3.2  Model verification using the computer code of the pioneer's publication

In this approach (**Figure 1**) the researcher has to look for two things (a) whether all the model specification statements in the publication text (**C**) can be found in **C*** and (b) whether all specification texts in **C*** have an analogue in **C**.  If the statement (a) is not true, i.e. there are **ΔC** statements in **C** that failed to find a place in **C*** the model fails verification. The model also fails verification if the statement (b) is not true, i.e. there are **ΔC\*** statements in **C\*** that failed to find a place in **C**.

A researcher with a good deal of familiarity with the programming environment of **I** can easily condense the logic underlying **I** into natural language statements that make up **C\***. However ratifying the equivalence of **C** and **C\*** is not easy.

We note that Judgements **J1** [**ΔC ≈ 0?**] and **J2** [**ΔC\* ≈ 0?**] in **Figure 1** depend heavily on the expertise of the researcher undertaking model verification. If the researcher fails to extend sufficient effort and/or gets fatigued without doing a thorough job a faulty model may be declared as having passed model verification. When can this happen? In other words, under what conditions will a researcher erroneously certify **ΔC** and **ΔC\*** are both zero, when at least one is not? I discuss this in the paragraphs that follow.

A publication text contains many statements that need not go into the computer code implementation. When one sets out to compare **C** with **C\***, familiarity with the implementation **I** – attained when developing **C\*** – tends to orient the researcher's mind regarding the components of the specification text that should go into the computer code implementation. As a result, some statements in **C** that do not find a place in **C\*** get ignored as noise. This carries the potential for misjudgment on **J1**.

Likewise, since computer code is much more detailed than natural language text specification, certain code sections in I – that got migrated to **C\*** – that do not have an easily-identifiable analogue in **C** get dismissed as *mere* implementation detail. This seeds the potential for misjudging **J2**.

Further, certain natural language sentences may be ambiguous, i.e. can suggest more than one computer code implementation. Here too, owing to prior familiarity with **I**, the researcher undertaking model verification is quite likely to accept the meaning espoused in the original computer code implementation, thereby failing to flag potential non-conformance. This becomes problematic when other researchers—who do not go through the computer code underlying an agent-based-model publication—assume a different meaning of the ambiguous sentence, and derive an entirely different meaning of the publication results.

In sum, contrary to popular belief, availability of the original computer code may lead to failure to effect model verification appropriately.

**Figure 2. Model verification by Conceptual Replication**

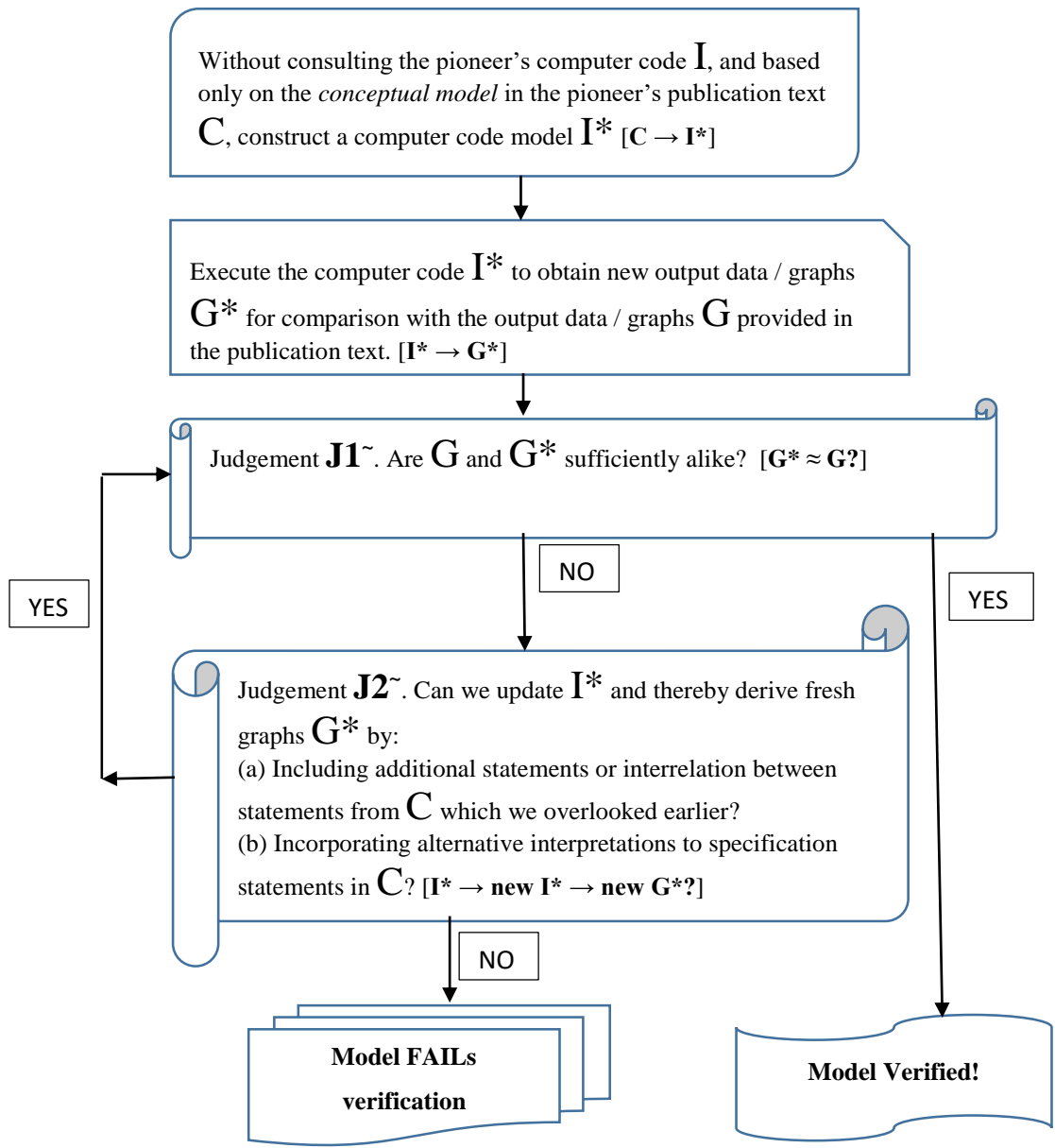

*Note.* This approach utilizes the pioneer's publication text **C** as well as the output graphs / data tables **G**. It does not require access to the pioneer's computer code **I**.

## 3.3  Model verification by Conceptual Replication

The second method of model verification, conceptual replication (**Figure 2**), implies that first, an independent computer code implementation (**I***) is developed from the text **C** of the pioneers' publication (without looking at **I**). Next, the graphical output **G***, obtained from **I***, is compared with the graphical output **G** (or table of outcome values, if graphs are absent) given in the publication text. If **G** and **G*** are sufficiently alike, the model passes verification.

Else, (i.e., upon failing to obtain sufficient match between **G** and **G\***), the publication text (**C**) is intensively scanned, repeatedly, in order to find specifications that got missed or misinterpreted when developing **I\***. If **G** and **G\*** are not sufficiently similar when further improvement to **I\*** is deemed impossible, the model fails verification.

In *conceptual replication*, the researcher undertaking model verification has to make judgments **J1~** and **J2~** repeatedly. Being wrong with respect to judgement **J1~** is next to impossible. On the other hand, if the researcher gives up on account of fatigue with respect to judgement **J2~** the model shall stay unverified. Thus, there is no question of a model erroneously obtaining a status of having been independently verified.

We can now articulate why *conceptual replication* is superior to the prior method (checking conformance between the text and code of the pioneer) for model verification. There are two reasons. First, *conceptual replication* requires deeper reflection on what computer statements appropriately represent **C**; starting from the pioneers' computer code (**I → C\***) does not. Second, persistence of mismatches between **G** and **G\*** trigger intensive scanning of the publication text (**C**), thereby helping remove variances between **C** and the independent computer code model **I\***. In the other method (checking conformance between the text and code of the pioneer) a similar compelling and persistent trigger is absent. Rather, statements in **C** not having an analogue in **I** tend to get dismissed as noise, and entire code segments in **I** may fail to come under scrutiny by being considered as *mere* implementation details.

In sum, conceptual replication is superior to the method of checking conformance between the text and code of the pioneer because the requirements for obtaining the status "*model verified*" are more demanding and on account of the fact that the pioneer's code is not necessary, given that the graphs/ output data tables in the pioneering publication are put to better use.

# 4   Discussion

The laudable goal of model verification may inadvertently get ill-served if journal editors give in to voices demanding compulsory public availability of computer code of computational simulation modeling / ABM publications. Adequate reflection on the pitfalls of such an action is called for. If computer code is publicly available, there is hardly any incentive to independently verify the mapping of the conceptual specifications in the original publication to the specifications for a computer code model. In this situation, inconsistencies in the pioneers' code are likely to persist in subsequent research, doing a major disservice to scientific progress. In order to encourage model verification, journals may consider publishing studies highlighting failed verification. The authors of the original study, if still active, may likewise be given an opportunity to file a response.

### 4.1 Limitations of this study

Critical evaluation of *appropriateness* of assumptions in an agent-based modeling study with respect to the physical processes of the phenomena being modeled is out of scope for this study. For example, in their critique of the garbage can model of decision making from Cohen *et al.* (1972), Bendor *et al.* (2001) point out that the setting of organized anarchy (an important contextual dimension of the *garbage can model*) is incompatible with an assumption that all decision makers follow the same heuristic to select decision areas to which they attend. As a second example, Edmonds and Hales (2003) persuasively argue for a far narrower scope than that claimed in the original study by Riolo *et al.* (2001). In the Riolo *et al.* (2001) mechanism, agents donate fitness to *only* those agents that have a similar tag (i.e., is similar with respect to one particular attribute value). It is seen that zones of cooperation come into being, even though each agent is otherwise self-interested. Edmonds and Hales (2003) show that the phenomenon of tag-based cooperation materialized only under compulsory donation to others, even if interacting agents have identical heritable tags, and only under a condition where when a given tag clone replicates itself exactly. In a third example, Chanda *et al.* (2018) show that the model of the continuum conception of exploration and exploitation (March 1991) was incomplete; bringing into consideration of a lower level of the stock variable (*collective human capital*) reversed an important result. These examples suggest that challenging the appropriateness of assumptions in an ABM requires not only a clear set of specification statements in the publication document, but also considerable domain knowledge.

Second, nothing prevents computational simulation modeling researchers from voluntarily making their computer code publicly available. This action may result in a certain degree of unreflective acceptance, stymieing model verification. However, given the prospect of a reward—space in a journal upon coming up with failed verification—the incentives to break the model (when it deserves to be broken) will be still present. In addition, availability of the pioneer's code can actually come to use to pinpoint the problematic sections in the pioneer's code (for an example see Chanda and Miller 2019).

### 4.2 Broader implications

Civilization progresses on the back of effective systems that (a) deploy human intelligence skillfully and (b) have feedback processes in place that appropriately go around human frailties. The point of view that insists on mandating public availability of computer code as a precondition to publication focuses on preventing cheating—researchers making untenable theoretical claims on the basis of compromised computer code. In the process it loses sight of the big picture: that the most important challenge is to develop a set of natural language specifications that suggest an equivalent set of computer instructions *across researchers*, as evidenced by congruence between numerical of graphical outputs from independently-developed computer code (*conceptual replication*).

Conceptual replication *directs* energy towards mental processes to derive computer instructions appropriate for a set of natural language arguments. The act of checking conformance with the numerical or graphical outputs of the original publication provides the right kind of *persistent* feedback towards more intensive and mindful efforts in scanning the specifications in the publication text.

Conversely, an attempt to check conformity between the publication text and the original computer code fails on both counts. Mental energy is not spent on the more difficult task of evolving computer code specifications from natural language statements. Rather, mental energy is spent on the trivial task of summarizing computer code blocks into natural language specifications. Moreover, the absence of persistent feedback encouraging more intensive consideration—i.e. *not* having to seek convergence of graphical outputs generated by two independently developed computer programs—creates conditions rife for giving in to human frailties. Sections of the publication text get dismissed as noise, and sections of the original computer code get dismissed as mere implementation detail.

The analysis above has important implications for design of human systems: political systems, the system of bureaucracy in a government, management systems in business organizations, and so forth. Frequently, we encounter a scenario where two organizations start with very similar endowments, but end up in very different places with respect to organizational success, after some time. The differing fortunes of the "Baby Bells" arising from the break-up of AT&T Bell Labs come to mind. It is quite likely that an organizational configuration will do better if skillful deployment of human intelligence is feasible because effort-directing feedback processes are in place to appropriately go around human the frailty of giving up in the absence of actionable feedback.

The study elegantly displays the working of Deutsch's *constructor theory* (Deutsch 2013). Constructor theory recommends articulating nature's laws in terms of what is possible and what isn't, and why so. Heisenberg's *uncertainty principle* is a classic example: a researcher cannot know *both* the position as well as the momentum of a sub-atomic particle (like electron), at any given point in time. To know the position the researcher will have to use light. Upon absorbing a photon from that incident light an electron goes to a higher energy state, i.e. momentum changes. In this case the researcher lacks a means other than throwing light, in order to assess the position of a particle. Alternately, tracks in a *Wilson Cloud Chamber* can provide information regarding the momentum of an electron, but the electron itself is long gone (i.e. position has changed). Thus, either a shortcoming of the inquiring researcher and/or limitation(s) of the tools available with the researcher *makes* something impossible.

Our study likewise suggests that it is impossible to verify a model by looking at the pioneer's code and publication text, except in simple and trivial cases. A researcher attempting to verify a complex model by merely looking at the pioneer's code and publication text will simply not have sufficient effort-directing feedback on his/her actions over time. Only when graphs / table of data from independently developed computer code are compared with that in the publication text, an appropriate guidepost (or beacon) is available to direct the researcher's activity along fruitful channels.

## 5 Conclusion

A belief that public availability of code underlying a computational simulation / agent-based-modeling publication is sufficient for model verification is misplaced. If journals need to choose between two policies (a) making public disclosure of computational simulation code compulsory OR (b) committing to provide space to failed replications, the second policy is preferable. This policy directs effort where it is most needed, i.e. towards verifying whether the set of natural language statements in the computational simulation modeling publication suggest a logically equivalent set of computer instructions across researchers.

**Author Biography**

**Sasanka Sekhar Chanda** is Professor in Strategic Management in Indian Institute of Management Indore. His research interest is in theory development by computational simulation modeling, in the domains of strategic decision-making, managerial intentionality, and complexity theory and study of project and system failures. Earlier, Sasanka worked in the industry in a range of roles spanning engineering, consulting, and management for a period of fifteen years**.** Some of his key publications are in *Computational and Mathematical Organization Theory*, *Decision*, *M@n@gement* and *Strategic Organization*.

# Appendix

## Prior model verification studies

I classify prior model verification studies into three categories: *docking*, *reimplementation* and *conceptual replication*. These labels have been suggested by researchers originating such studies. The main point of difference between the categories is with respect to the goal of the study. Below, I elaborate on these stylized types by an example each.

### Docking

In some contexts, it is desirable to verify whether two different models—that incorporate distinct, alternative theories—produce similar pattern of outputs. Modifying one model to produce outputs similar to the other is what lies at the heart of *docking*. Axtell *et al.* (1996) describe an initiative in which the *Sugarscape* model by Epstein and Axtell (1996) was updated with some assumptions from the cultural model by Axelrod (1995, 1997). The objective was to check if outcomes from the two models would be similar. Both these are agent based models. In the *Sugarscape* model agents have behavioral and cultural rules, and can move around in a space with a toroidal topology—where every agent has the same number of neighbors. In the Axelrod cultural model, agents are stationary and placed randomly in a square lattice. In each period an agent randomly contacts a neighbor, to acquire a (random) cultural attribute. Transmission is allowed only if there is at least one attribute in which the interacting pair of agents does not differ. Simulations stop when every agent has attributes identical to its neighbors, or has no attribute in common with its neighbors.

To carry out docking, the *Sugarscape* model was updated to have stationary agents on a square lattice. Further the agents were configured to be activated at random order instead of in a random sequence[2]. It was found that *Sugarscape* could generate output patterns comparable to that of the Axelrod cultural model leading to a common inference—that cultural change becomes more likely as two neighbors are more alike, and less likely as they differ. The Axelrod cultural model demonstrated the formation of multiple cultural configurations at equilibrium. Axtell *et al.* (1996) further show that, when movement of agents is activated (i.e. spatial correlation is reduced), convergence to equilibrium occurs later, and there occur lower number of cultures at equilibrium. In sum, the exercise of docking showed that the assumption concerning high spatial correlation (stationary agents) was an important contributor to proliferation multiple cultures at equilibrium.

---

[2] In the original *Sugarscape* model, each agent would be activated exactly once in a cycle.

**Reimplementation**

. In *reimplementation*, a model that is a reproduction of a prior model differs from the latter in one or more from six dimensions: time, hardware, computer language used, toolkits, algorithms and authors (Wilensky and Rand 2007). Keeping the other five dimensions constant, if a model is attempted to be recreated at a later point in time, the possibility of obtaining a variance is the least if the conceptual model described in the text of the prior publication is accurate. If such an issue is absent, further changing the computer language will normally yield no variance unless some model result depends crucially on variation in floating point arithmetic and/or on inbuilt process for generating random numbers—mechanisms that are likely to be somewhat different across computer languages.

A toolkit is a set of program libraries written in a particular language (Wilensky and Rand 2007). Failure of a reproduction attempt by a toolkit would, in most cases, be an evidence of problems with the toolkit itself. In such a scenario, the later researcher should attempt to reproduce a program without assistance of toolkits. Alternately, if the original computer program was written using toolkits, variation of its output from that of a program written without assistance from a toolkit reflects inadequate understanding of the toolkit's algorithm on the part of the follow-on researcher, or logical issues in the design of the toolkit with which the original program was written.

Differences in program output arising from differences in algorithm are of the type discussed under conceptual replication above. Lastly, if different authors build models that give different pattern of output, it is a sign that the text of the original publication may be ambiguous in some places. Alternate interpretations of the same text lead to outcome variance.

Wilensky and Rand (2007) describe how they reproduced the model in Axelrod and Hammond (2003, referred in Wilensky and Rand 2007) in *NetLogo*, a software platform different from that of the original study. Initially their outcomes were different from that of the original. However, when some of the finer details of the Axelrod and Hammond (2003) model were incorporated, they were able to get the results nearly identical to the original.

**Conceptual replication**

In *conceptual replication*, a follow-on researcher consults the model specification in the publication text in order to independently develop computer code, and compares outputs with that in the publication. In case certain important parameters or mechanisms are missing from the publication text, "inconsistencies in outputs between the two implementations" (Edmonds and Hales 2003, p. 2) get revealed.

Chanda and Miller (2019) describe their experience of failing to replicate the graphical output of March (1991), a highly-cited genetic algorithm[3] model illustrating exploration and exploitation[4] by developing program code solely from the publication text. Upon obtaining the original program code from Prof. March they found that there were three additional features in that code that were either not present in the publication text, or at variance with specifications in the publication text. Fortunately though, March's theoretical findings actually got accentuated after removal of the undocumented features, paving the way for deployment of this model in further theory-development studies in social science (e.g. Chanda and Ray 2015; Chanda 2017; Chanda and McKelvey 2020).

**Synthesis of learning from prior model verification studies**

The instances discussed above illustrate that inadequate details regarding the computer code implementation in the original publication text constitute the most likely reason for replication failure. A related and close second reason is ambiguity in the publication text: in such cases, there exist multiple interpretations for a natural language statement, and a later researcher cannot easily determine what meaning the pioneers may have implemented. The case whereby computer code implements features that go unmentioned in the publication text constitute the third ground for difficulty with replication. In all these cases the problems surface when a later researcher attempts replication of published results without consulting the pioneers' code. The problems get recognized as problems when the later researcher fails to obtain graphical (or numerical) characteristics matching that of the pioneers. Indeed the mismatch serves to direct effort till either convergence is accomplished, or the model is definitely known to have failed verification.

**Additional References**[5]

---

[3] March, in turn draws from Holland's (1975) work.
[4] Exploration involves experimenting with heterogeneous knowledge, with the objective of developing new products and services; exploitation involves doing known things better, through refinement (Chanda and McKelvey, 2020)
[5] References common to main document are color tagged.